\documentclass[prb,showpacs,aps,twocolumn,A4paper,floatfix,superscriptaddress]{revtex4}

\usepackage[dvips]{graphics}
\usepackage{color}
\usepackage{epsfig}
\usepackage{xspace}
\usepackage{amssymb,amsmath}

\def\1m10{[1$\bar{1}$0]\mbox{ }}
\def\Schar1m10{\{1$\bar{1}$0\}\mbox{ }}
\def\1m102{[1$\bar{1}$0]}

\def\p4x4{$p(4 \times 4)$\ }

\def\c4x8{$c(4 \times 8)$\ }

\begin{document}

\title{Experimental and theoretical study of oxygen adsorption structures on Ag(111)}

\author{Joachim~Schnadt}
\email{joachim.schnadt@sljus.lu.se}
\affiliation{Division of Synchrotron Radiation Research, Department of Physics, Lund University, Box 118, 221 00 Lund, Sweden}
\affiliation{Interdisciplinary Nanoscience Center (iNANO) and Department
of Physics and Astronomy, University of Aarhus, Building 1521, Ny Munkegade, 8000 Aarhus C, Denmark}
\author{Jan Knudsen}
\affiliation{Interdisciplinary Nanoscience Center (iNANO) and Department
of Physics and Astronomy, University of Aarhus, Building 1521, Ny Munkegade, 8000 Aarhus C, Denmark}
\author{Xiao~Liang~Hu}
\affiliation{London Centre for
Nanotechnology and Department of Chemistry, University College
London, London WC1E 6BT, UK}
\author{Angelos~Michaelides}
\affiliation{London Centre for
Nanotechnology and Department of Chemistry, University College
London, London WC1E 6BT, UK}
\affiliation{Fritz-Haber-Institut der Max-Planck-Gesellschaft, Faradayweg 4-6, 14195 Berlin, Germany}
\author{Ronnie T. Vang}
\affiliation{Interdisciplinary Nanoscience Center (iNANO) and Department
of Physics and Astronomy, University of Aarhus, Building 1521, Ny Munkegade, 8000 Aarhus C, Denmark}
\author{Karsten~Reuter}
\affiliation{Fritz-Haber-Institut der Max-Planck-Gesellschaft, Faradayweg 4-6, 14195 Berlin, Germany}
\author{Zheshen Li}
\affiliation{Institute for Storage Ring Facilities, University of
Aarhus, Building 1525, Ny Munkegade, 8000 Aarhus C, Denmark}
\author{Erik L{\ae}gsgaard}
\affiliation{Interdisciplinary Nanoscience Center (iNANO) and Department
of Physics and Astronomy, University of Aarhus, Building 1521, Ny Munkegade, 8000 Aarhus C, Denmark}
\author{Matthias~Scheffler}
\affiliation{Fritz-Haber-Institut der Max-Planck-Gesellschaft, Faradayweg 4-6, 14195 Berlin, Germany}
\author{Flemming Besenbacher}
\email{fbe@inano.dk}
\affiliation{Interdisciplinary Nanoscience Center (iNANO) and Department
of Physics and Astronomy, University of Aarhus, Building 1521, Ny Munkegade, 8000 Aarhus C, Denmark}

\date{\today}

\begin{abstract}
    The oxidized Ag(111) surface has been studied by a combination of experimental and theoretical methods,
    scanning tunneling microscopy (STM), x-ray photoelectron spectroscopy (XPS), and density functional theory (DFT).
    A large variety of different surface structures is found, depending on the detailed preparation conditions.
    The observed structures fall into four classes: (a) individually chemisorbed atomic oxygen atoms,
    (b) three different oxygen overlayer structures, including the well-known $p(4 \times 4)$ phase,
    formed from the same Ag$_6$ and Ag$_{10}$ building blocks, (c) a {$c(4 \times 8)$} structure not previously
    observed, and (d) at higher
    oxygen coverages structures characterized by stripes along the high-symmetry directions of the Ag(111) substrate. Our analysis
    provides a detailed explanation of the atomic-scale geometry of the Ag$_6$/Ag$_{10}$ building block structures,
    and the {$c(4 \times 8)$} and stripe structures are discussed in detail. The
    observation of many different and co-existing structures implies that the O/Ag(111) system is
    characterized by a significantly larger degree of complexity than previously anticipated, and this will impact our
    understanding of oxidation catalysis processes on Ag catalysts.
\end{abstract}


\maketitle

\section{Introduction}

\begin{figure*}
    \begin{center}
        \includegraphics[width=17.8cm]{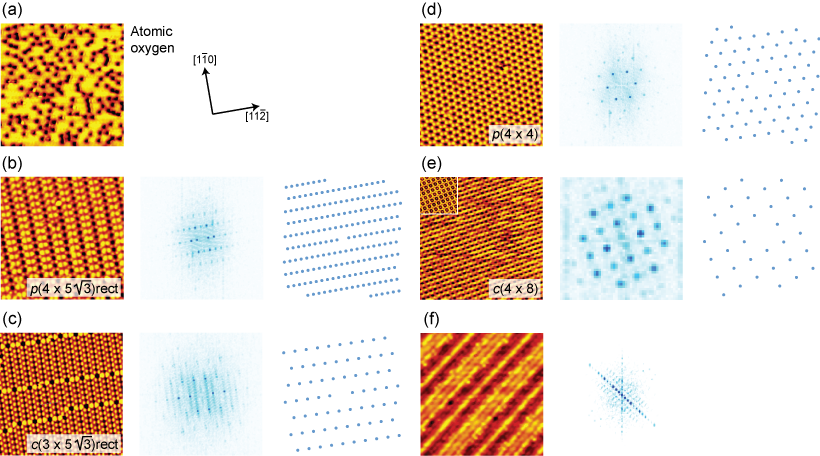}
    \end{center}
    \caption{\label{FigAllStructures}(Color online) Phases of oxidized Ag(111) observed in the present
    study. Further phases observed previously are specified in
    Table~\protect\ref{TableConditions}. The STM images to the
    left in the panels are supplemented by the FFTs of the STM
    images (middle parts of the panels) and by the theoretical
    LEED images simulated on the basis of the symmetry
    assignements (right-hand parts).
    The surface directions for all images are indicated in panel (a).
    (a) $70 \mbox{ \AA} \times 70 \mbox{ \AA}$, 1.12~nA, 9.8~mV.
        The depressions
        are interpreted as chemisorbed atomic oxygen with a coverage of approximately 11\% of a monolayer.
    (b) $200 \mbox{ \AA} \times 200 \mbox{ \AA}$, 0.62~nA, 39.1~mV.
        $p(4 \times 5 \sqrt{3})$rect.
    (c) $200 \mbox{ \AA} \times 200 \mbox{ \AA}$, -1.2~nA, -92.8~mV.
        $c(3 \times 5 \sqrt{3})$rect.
    (d) $200 \mbox{ \AA} \times 200 \mbox{ \AA}$, 0.99~nA, 39.1~mV.
        $p(4 \times 4)$.
    (e) $200 \mbox{ \AA} \times 200 \mbox{ \AA}$, 0.64~nA, 110.5~mV.
        $c(4 \times 8)$.
        The inset shows the same surface phase measured with a different tip state.
        $60 \mbox{ \AA} \times 60 \mbox{ \AA}$,
        -0.34~nA, -109.6~mV.
    (f) $200 \mbox{ \AA} \times 200 \mbox{ \AA}$, 0.41~nA, 1250,0~mV.
        Stripe structure.}
\end{figure*}

\begin{figure}
    \begin{center}
        \includegraphics[width=8.6cm]{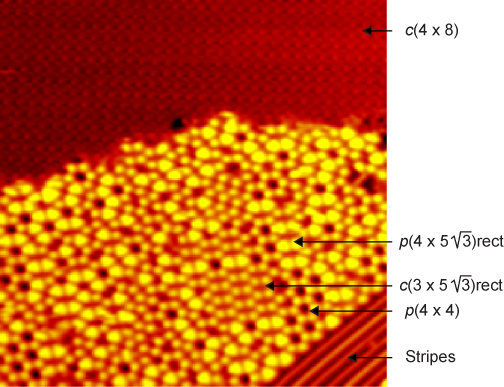}
    \end{center}
    \caption{\label{FigStructuresAtthesametime}(Color online)
    Coexistence of phases.
    $250 \mbox{ \AA} \times 250 \mbox{ \AA}$ STM image of a Ag(111) sample exposed to $5 \times 10^{-8}$~Torr
   atomic oxygen for 40 min at 500 K sample temperature (-0.44~nA, -55.8~mV).}
\end{figure}

The oxidation of the Ag(111) surface is a fascinating example of
how a seemingly simple system can escape a detailed understanding
for decades, in spite of repeated and thorough efforts. The
history of surface science investigations of the oxidation of
Ag(111), briefly reviewed recently,\cite{Michaelides2005} started
back in the sixties and early seventies with a couple of single
crystal studies\cite{Mueller1966,Degeilh1969,Dweydari1973} and
took a serious upswing with the investigations by Rovida
\textit{et al.} in the seventies,\cite{Rovida1972,Rovida1974} who,
in particular, reported on the now renowned $p(4 \times 4)$ phase.
In the following much of the effort was directed towards this
phase,\cite{Albers1977,Campbell1985,Bare1995,Bukhtiyarov1999,Carlisle2000a,Carlisle2000b,Li2003,Li2003a,Li2003b,Michaelides2003,Bocquet2003a,Bocquet2003b,Michaelides2005,Schnadt2006,Schmid2006,Reichelt2007}
but also numerous other studies of more general character
appeared,\cite{Engelhardt1976,Albers1977a,Albers1977,Benndorf1983,Grant1984,Campbell1985,Schmeisser1985,Campbell1986,Spruit1989,Spruit1989a,Wang1991,Reijnen1991,Bao1993,Pettinger1994,BuatierdeMongeot1995,Raukema1996,Bao1996,Lacombe1996,Rocca1997,SchedelNiedrig1997,Bukhtiyarov1999,Li2002,Li2003,Michaelides2003,Stegelmann2004,Reichelt2007,Reicho2007,Reichelt2007a}
as well as works concerned with the reactivity of O/Ag(111)
towards, in particular, the oxidation of ethene and methanol.
\cite{Campbell1985b,Campbell1985c,Tan1987,vanSanten1987,Hawker1989,Mukoid1990,Wu1994,Carley1998,Boronin1999,Avdeev2001,Scheer2002,Stacchiola2001,Klust2006,Gao2007,Gomes2007,Greeley2007,Zhou2008,Huang2008,Christopher2008}
Only recently the atomic-scale geometric model for the $p(4 \times
4)$ phase was completely revised on the basis of a combination of
scanning tunnelling microscopy (STM), surface x-ray diffraction
(SXRD), and x-ray photoemission spectroscopy (XPS) experiments and
density functional theory
calculations.\cite{Schnadt2006,Schmid2006} Subsequent, detailed
low-energy electron diffraction and CO reaction studies supported
this atomic model.\cite{Reichelt2007a,Klust2007} At the same time
an increasing number of studies have indicated that the phase
diagram of the oxidized Ag(111) surface is more complex than
previously anticipated. While earlier studies only stipulated the
existence of the $p(4 \times 4)$ and an atomic adsorbate layer
(including a phase with a local
$p(\sqrt{3}\times\sqrt{3})$R$30^\circ$ symmetry), recent studies
have reported phases containing both less\cite{Reicho2007},
equally much,\cite{Carlisle2000b,Schnadt2006,Reichelt2007} and
more\cite{Schnadt2006} oxygen than the $p(4 \times 4)$ phase (cf.
Table~\ref{TableConditions}). Many of the reactivity studies have
assumed that the catalytical properties of Ag are associated with
the $p(4 \times 4)$ structure due to the anticipation that this
phase should be predominant under oxygen pressures similar to
those under reaction conditions.\cite{Li2003b,Li2003a} These
investigations have in general not taken into account the
existence of further phases and, hence, a much more complex phase
diagram for the Ag-O system will fundamentally change our picture
of what happens in the catalytic process. At present, growing
evidence exists that the static phases found in surface science
studies may neither be responsible for the surface's catalytic
activity itself nor be preserved during catalytic
operation.\cite{Stampfl2002,Hendriksen2004,Ackermann2005,Reuter2006,Reichelt2007}
The surface is rather to be considered as a dynamic medium, the
structure of which changes in response to the changing chemical
environment. The formation and breakup of these static Ag-O phases
are still of significant interest due to the phases' role in the
overall dynamics of the catalytic process. Thus surface science
studies with their unprecedented ability of clarifying the
atomic-scale structure still retain their value and validity.

Here we present a detailed and thorough experimental and
theoretical study of oxygen adsorption on Ag(111), with data
concerning both previously observed structures and hitherto new
unobserved phases of oxidized Ag(111). The present study shows
that the $p(4 \times 4)$ phase is embedded in a wider context of
two additional phases, the $c(3 \times 5 \sqrt{3})$rect and $p(4
\times 5 \sqrt{3})$rect structures. It is also found that the $p(4
\times 4)$ structure, to our knowledge in contrast to the $c(3
\times 5 \sqrt{3})$rect and $p(4 \times 5 \sqrt{3})$rect phases,
can host foreign atoms and molecules. Finally, the existence of
two more phases with a $c(4 \times 8)$ and "stripe" character is
reported.

\section{Experimental and theoretical methods}

The experiments were carried out in the STM laboratory and at the
vacuum ultraviolet/soft x-ray synchrotron radiation facility
ASTRID at Aarhus University. In the STM laboratory we used an
ultra-high vacuum (UHV) chamber with a base pressure of $1 \times
10^{-10}$ Torr, equipped with standard instrumentation for sample
cleaning, the Aarhus STM,\cite{Laegsgaard1988} a low-energy
electron diffractometer, and a thermal gas cracker from Oxford
Instruments for atomic oxygen exposure with a cracking efficiency
which in our experiments varied from 30\% at $10^{-9}$ to 14\% at
$10^{-6}$ Torr total pressure. Directly connected to the UHV
chamber is a small and compact high pressure cell which makes it
possible to dose molecular oxygen at pressures up to one
atmosphere. After oxygen exposure the sample was characterized by
STM and, in some cases, low-energy electron diffraction (LEED).
The transfer between the high pressure cell and the UHV chamber
could be accomplished within less than 15 min of oxygen exposure.
Most of the STM experiments were performed at room temperature,
although some of the STM images were recorded at liquid nitrogen
temperature, which sometimes resulted in higher spatial
resolution.

For the XPS measurements the surfaces were prepared and initially
characterized in the STM chamber and then brought to the SX-700
beamline\cite{Uggerhoej1995} of ASTRID in a vacuum suitcase with a
base pressure better than $1 \times 10^{-9}$ Torr. The pressure
during transfer from the vacuum chambers to the suitcase, which
could be accomplished within a minute, rose to in between $1
\times 10^{-8}$ and $1 \times 10^{-7}$ Torr. The base pressure of
the UHV chamber of the SX-700 beamline is $1 \times 10^{-10}$
Torr. After the XPS measurement the sample was transferred back to
the STM chamber and the status of the sample reinvestigated by STM.
The control experiments ensured that the examined structures were still present
and clearly visible in the STM, although some minor deterioration of
the surface might have occurred.

The Ag(111) surface was cleaned by repeated sputter/annealing
cycles. It was then oxidized by exposing it to either atomic
oxygen at partial atomic oxygen pressures of $1 \times 10^{-9}$ to
$1 \times 10^{-6}$ Torr or molecular oxygen at pressures between
0.5 and 10~Torr. During oxygen exposure the  Ag(111) sample was
held at temperatures in between 420 and 600~K for an exposure time
of 10 to 50~min. Following this recipe, the structure of the
oxidized surface varied with oxygen pressure, sample temperature,
and exposure time. As discussed below, the prepared surface
typically exhibited co-existing domains of different surface
phases. However, it was possible to prepare single phase surfaces
for some of the structures presented below by carefully tuning the
preparation parameters.

Bare \textit{et al.},\cite{Bare1995} have proposed an alternative
preparation procedure, in which the surface is exposed to NO$_2$
at pressures between $5 \times 10^{-8}$ to $5 \times
10^{-6}$~Torr, while the sample is kept at the same temperatures
as described above. This method has also been used in a variety of
subsequent
studies.\cite{Carlisle2000a,Carlisle2000b,Scheer2002,Huang2002,Huang2002a,Bocquet2003,Huang2003,Webb2004,Alemozafar2005,Reichelt2007a,Klust2007,Zhou2008a}
Here we do not present any results obtained on samples prepared by
this NO$_2$ method, although we did carry out a limited number of
such studies. Consistent with Carlisle \textit{et
al.},\cite{Carlisle2000b} we were able to produce the atomic
oxygen, $p(4 \times 5 \sqrt{3})$rect-O, and $p(4 \times 4)$-O
phases in these studies, which indicates that the use of NO$_2$
instead of atomic or molecular oxygen does not significantly
change the picture drawn up here.

All the density functional theory (DFT) calculations reported here
were carried out with the Perdew-Burke-Ernzerhof generalized
gradient approximation\cite{Perdew1996} in periodic supercells
within the plane-wave pseudopotential formalism as implemented in
the CASTEP code.\cite{Segall2002} As discussed below, a variety of
unit cells were used, all consisting of five layer Ag(111) slabs
and with fine Monkhorst-Pack \textit{\textbf{k}}-point meshes
equivalent to at least 24$\times$24$\times$1 per (1$\times$1)
surface unit cell. During structure optimizations the top layer of
Ag atoms as well as the oxide overlayer atoms were allowed to
relax, whilst the bottom four layers of Ag were held fixed. From
the DFT results simulated STM images were obtained using the
simple Tersoff-Hamann approximation.\cite{Tersoff1985} A ``tip''
height of 2.5 \AA\ above the highest atom in each overlayer for
occupied states within 2 eV of the Fermi level was used.

\section{Results and discussion}

\subsection{Overview of the observed surface structures}

Figure~\ref{FigAllStructures} displays STM images of all the
phases which we have observed experimentally. The symmetry
assignments ($c(3 \times 5 \sqrt{3})$rect, $p(4 \times 5
\sqrt{3})$rect, and $c(4 \times 8)$ for the hitherto non-assigned
phases; see below for a more detailed description) were derived on
the basis of the STM results. The middle parts of the panels
represent the fast Fourier transform (FFT) of the STM images,
while the right-hand parts reproduce LEED simulations obtained
from the symmetry assignments provided on the basis of the STM
observations. These LEED results are in excellent agreement with
the FFTs, which lends further credibility to the STM-based
symmetry assignments for the oxygen-induced phases.

At the lowest oxygen coverages an apparently disordered
arrangement of depressions is observed in the constant current STM
images (Fig.~\ref{FigAllStructures}(a)). The depressions are
interpreted as atomic oxygen adsorbates, consistent with previous
studies\cite{Carlisle2000b} and the well-known fact that oxygen
depletes the local density of states at the Fermi level, leading
to a depression-like appearance of atomic oxygen on metal
surfaces.\cite{Besenbacher1996} Domains of this disordered
structure are also frequently observed to co-exist with domains of
the structures shown in Figs.~\ref{FigAllStructures}(b) to (d).

\begin{table}
    \caption{\label{TableConditions}Summary of the structures, expected oxygen coverage, and preparation conditions
    for oxygen on Ag(111). For each preparation the first unequivocal
    identification (albeit not necessarily with the correct symmetry assignment and geometry)
    in literature is given. The asterisk indicates that the structure has been observed in the
    present study.}
    \begin{center}
    \begin{ruledtabular}
    \begin{tabular}[t]{ccccc}
        Structure & Oxygen   & Atomic & Molecular & NO$_2$ \\
                  & coverage & oxygen & oxygen    &        \\
        \hline
        Atomic oxygen & $\leq$ 0.05 ML\footnotemark[1] & $\times$\footnotemark[2]$^{,\ast}$
                      & $\times$\footnotemark[3]$^{,\ast}$
                      & $\times$\footnotemark[1]$^{,\ast}$ \\
        $p(\sqrt{3} \times \sqrt{3})$R30${^\circ}$ & 0.33 ML\footnotemark[4] &
                                               & $\times$\footnotemark[5]
                                               &                           \\
        $p(4 \times 5 \sqrt{3})$rect & 0.375 ML & $\times$\footnotemark[6]$^{,\ast}$
                                    & $\times$\footnotemark[3]$^{,\ast}$
                                    & $\times$\footnotemark[1]$^{,\ast}$ \\
        $c(3 \times 5 \sqrt{3})$rect & 0.4 ML & $\times$\footnotemark[6]$^{,\ast}$
                                     & $\times$\footnotemark[2]$^{,\ast}$
                                     & \\
        $p(4 \times 4)$ & 0.375 ML & $\times$\footnotemark[6]$^{,\ast}$
                        & $\times$\footnotemark[7]$^{,\ast}$
                        & $\times$\footnotemark[8]$^{,\ast}$ \\
        $p(7 \times 7)$ & 0.37 ML\footnotemark[9] &
                        & $\times$\footnotemark[10]
                        & \\
        $c(4 \times 8)$ & 0.25 ML\footnotemark[11] & $\times$\footnotemark[2]$^{,\ast}$
                        &
                        & \\
        Stripe phase    & ? &  $\times$\footnotemark[2]$^{,\ast}$
                          & $\times$\footnotemark[2]$^{,\ast}$
                          & \\
    \end{tabular}
    \end{ruledtabular}
    \footnotetext[1]{Reference \onlinecite{Carlisle2000b}.}
    \footnotetext[2]{Present work.}
    \footnotetext[3]{Reference \onlinecite{Reichelt2007}.}
    \footnotetext[4]{Based on the model in Ref.
    \onlinecite{Bao1993}.}
    \footnotetext[5]{A $p(\sqrt{3} \times \sqrt{3})$R30${^\circ}$ overlayer
    was identified by M{\"{u}}ller\cite{Mueller1966} on epitaxially grown
    Ag(111) films for relatively small oxygen exposures (100 L).
    Later results on a Ag(111) single crystal, which show no formation
    of overlayers after oxygen exposure up to 3000 L under
    otherwise similar conditions, contradict these initial
    experiments.\cite{Engelhardt1976} An overlayer with a local
    $p(\sqrt{3} \times \sqrt{3})$R30${^\circ}$
    symmetry was once again reported by Bao \textit{et al.} who used
    STM to study Ag samples exposed to pressures around one
    atmosphere for long times on the order of days\cite{Bao1993,Bao1996}
    (see also Ref.~\onlinecite{Schubert1995}). Overall, the observed phase
    was argued to be of higher-order commensurability.}
    \footnotetext[6]{Reference \onlinecite{Schnadt2006}.}
    \footnotetext[7]{Reference \onlinecite{Rovida1972}.}
    \footnotetext[8]{Reference \onlinecite{Bare1995}.}
    \footnotetext[9]{Based on the model in Ref. \onlinecite{Reicho2007}.}
    \footnotetext[10]{Reference \onlinecite{Reicho2007}.}
    \footnotetext[11]{Based on the tentative model in Section~\ref{Secc4x8stripe}.}
    \end{center}
\end{table}

\begin{table*}
    \caption{\label{TableSinglePhases}Summary of the preparation conditions, for which surfaces were obtained
    characterized by the predominance of a single phase.}
    \begin{center}
    \begin{ruledtabular}
    \begin{tabular}[t]{ccccc}
        Structure & Oxidizing agent & Pressure & Sample temperature & Exposure time \\
        \hline
        $p(4 \times 5 \sqrt{3})$rect                 & Atomic O & $5 \times 10^{-9}$ Torr & 500 K       & 10 min  \\
        $p(4 \times 5 \sqrt{3})$rect + bare Ag(111)  & Atomic O & $1 \times 10^{-8}$ Torr & 450 - 500 K & 5 - 15 min  \\
        $c(3 \times 5 \sqrt{3})$rect\footnotemark[1] & Atomic O & $1 \times 10^{-7}$ Torr & 500 K       & 10 - 20 min  \\
        $p(4 \times 4)$                              & O$_2$    & 10 Torr                 & 500 K       & 10 min \\
        Stripes                                      & Atomic O & $1 \times 10^{-6}$ Torr & 500 K       & 10 min \\
    \end{tabular}
    \end{ruledtabular}
    \footnotetext[1]{Contained "defects" of the $p(4 \times 5 \sqrt{3})$rect phase (cf. Figure~\ref{FigAllStructures}(c)).}
    \end{center}
\end{table*}

The structure in Fig.~\ref{FigAllStructures}(b) was previously
observed by Carlisle \textit{et al.}\cite{Carlisle2000b} and was
assigned to the decomposition of the $p(4 \times 4)$ structure in
panel (d). The structure, which we briefly mentioned in a previous
report,\cite{Schnadt2006} has a $p(4 \times 5 \sqrt{3})$rect unit
cell in the convenient terminology for rectangular symmetries on
hexagonal surfaces introduced by Biberian and van
Hove.\cite{Biberian1984} In matrix notation the unit cell (the
conventional and primitive unit cells are identical) is described
by
\begin{displaymath}
    \left(
    \begin{array}{rr}
        4 & 0 \\
        5 & 10
    \end{array}
    \right).
\end{displaymath}

The structure in Fig.~\ref{FigAllStructures}(c) also possesses a
rectangular symmetry with a conventional unit cell of $c(3 \times
5 \sqrt{3})$rect symmetry. The brighter lines from left to right
are units of the $p(4 \times 5 \sqrt{3})$rect structure. We have
observed that domains of the $c(3 \times 5 \sqrt{3})$rect
structure always contain at least small fractions of defects of
$p(4 \times 5 \sqrt{3})$rect symmetry, which in room temperature
experiments sometimes have been observed to move along the \1m102
direction. The primitive unit cell of the $c(3 \times 5
\sqrt{3})$rect phase is given by\cite{PreviousNotation}
\begin{displaymath}
    \left(
    \begin{array}{rr}
        3 & 0 \\
        4 & 5
    \end{array}
    \right).
\end{displaymath}

Panel (d) of Fig.~\ref{FigAllStructures} shows the well-known $p(4
\times 4)$ structure, which has been the subject of surface
science investigations since the early 1970s, whereas the
structures in panels (e) and (f) are reported here for the first time.
The first of these has a $c(4 \times 8)$ symmetry, while the
appearance of the second structure is characterized by stripes of
variable thickness along the \Schar1m10 surface directions. In
matrix notation the primitive unit cell of the $c(4 \times 8)$
structure is given by
\begin{displaymath}
    \left(
    \begin{array}{rr}
        3 & 1 \\
        1 & 3
    \end{array}
    \right).
\end{displaymath}

Table~\ref{TableConditions} summarizes the conditions, which here and
in other studies have been used to prepare the different surface
structures. Both the atomic and molecular oxygen recipes are very
versatile in terms of the variety of structures which can be
produced. The NO$_2$ recipe has so far primarily been used to
prepare oxidized Ag(111) surfaces covered entirely by
the $p(4
\times 4)$
structure.\cite{Bare1995,Carlisle2000a,Carlisle2000b,Schmid2006,Reichelt2007a}
The present results show that also the $p(4 \times 5
\sqrt{3})$rect structure can be produced by exposing the Ag(111) crystal to NO$_2$. Due to the
similarity of the $p(4 \times 4)$, $p(4 \times 5 \sqrt{3})$rect,
and $c(3 \times 5 \sqrt{3})$rect structures (see below) it can be
expected that it also is possible to form the latter using NO$_2$,
while it is more difficult to draw conclusions along these lines
regarding the remaining structures.

For most preparation conditions a variety of co-existing surface
structures appear. Only for the preparation conditions summarized
in Table~\ref{TableSinglePhases} we were able to obtain
single-phase or close-to single-phase surfaces. The otherwise
observed co-existence of different surfaces structures is
illustrated in Fig.~\ref{FigStructuresAtthesametime}, which
displays an STM image recorded after exposing the surface to $5
\times 10^{-8}$~Torr atomic oxygen at 500~K for 40~min. In this
STM image the $p(4 \times 5 \sqrt{3})$rect, $c(3 \times 5
\sqrt{3})$rect, $p(4 \times 4)$, $c(4 \times 8)$, and stripe
structures are found to co-exist within a 250~{\AA} $\times$
250~{\AA} area of the surface. It is observed that the $p(4 \times
5 \sqrt{3})$rect, $c(3 \times 5 \sqrt{3})$rect, and $p(4 \times
4)$ structures have very similar STM appearances with respect to
their height profiles (brightness), whereas the STM height profile
for the the $c(4 \times 8)$ structure is much lower. As will be
seen and discussed in further detail below this reflects that the
atomic geometries of the $p(4 \times 5 \sqrt{3})$rect, $c(3 \times
5 \sqrt{3})$rect, and $p(4 \times 4)$ structures are indeed very
similar.

\begin{figure*}
    \begin{center}
        \includegraphics[width=17.8cm]{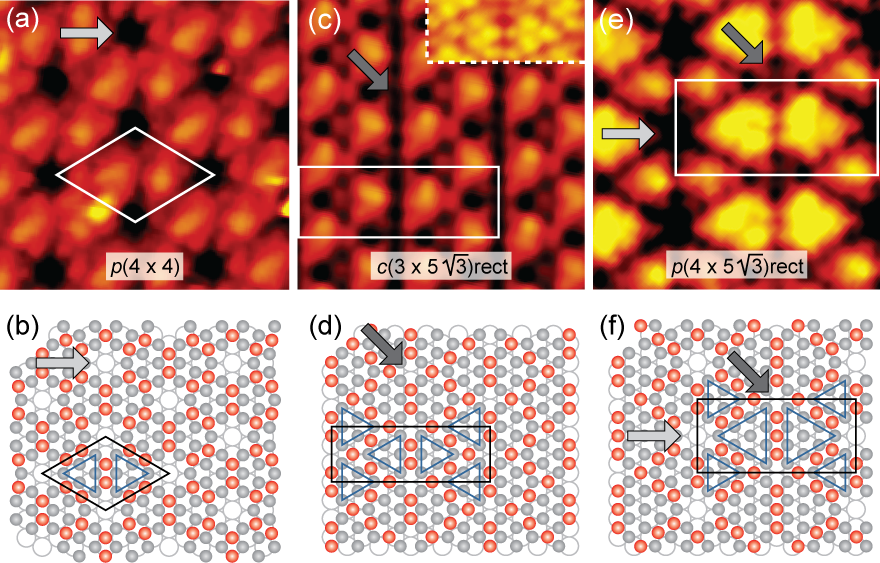}
    \end{center}
    \caption{\label{FigBuildingBlockStructures}(Color online)
    (a), (c), (e) STM images and (b), (d), (f) experimentally derived structures of the Ag$_6$/Ag$_{10}$ building block structures.
    The structural models for the $p(4 \times 4)$ and $c(3 \times 5 \sqrt{3})$rect phases
    in panels (b) and (d) also contain the in
    Ref.~\protect\onlinecite{Schnadt2006} theoretically derived
    positions of the oxygen atoms.
    The red/dark gray spheres represent the oxygen, (medium)
    gray the overlayer silver, and the open spheres the
    Ag(111) substrate atoms. The blue lines are guides for the eye indicating the
    Ag$_6$ and Ag$_{10}$ building blocks.
    (a) $35 \mbox{ \AA} \times 35 \mbox{ \AA}$, -0.42~nA, -21.7~mV. $p(4 \times 4)$.
    (b) $35 \mbox{ \AA} \times 35 \mbox{ \AA}$, -0.40~nA, -34.2~mV. $c(3 \times 5 \sqrt{3})$rect.
    (c) $35 \mbox{ \AA} \times 35 \mbox{ \AA}$, -0.38~nA, -17.4~mV. $(4 \times 5 \sqrt{3})$rect.
    The inset in panel (c) shows a highly resolved Ag$_6$ dimer with all 12 Ag atoms visible (-0.42~nA, -21.7~mV).
    The light (dark) arrows mark the voids characteristic for the $p(4 \times 4)$ ($c(3 \times 5 \sqrt{3})$rect-like)
    phase. Both types of voids are found in the
    $p(4 \times 5 \sqrt{3})$rect structure.}
\end{figure*}

\begin{figure}
    \begin{center}
        \includegraphics[width=8.6cm]{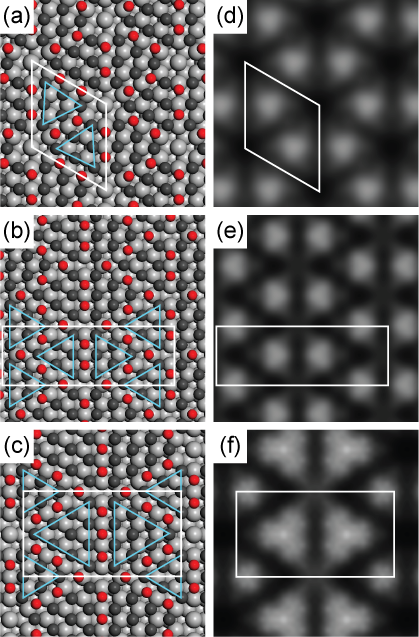}
    \end{center}
    \caption{\label{FigTheoryBBStructures}(Color online) (a) to (c) Final geometrical models of the
    building block structures $p(4 \times 4)$, $c(3 \times 5 \sqrt{3})$rect, and
    $p(4 \times 5 \sqrt{3})$rect. (d) to (f) Tersoff-Hamann simulations of the structures in (a) to (c).
    The dark (light) gray spheres represent Ag atoms in the overlayer (substrate) and the smaller red spheres O atoms.
    The blue lines are guides for the eye indicating the
    Ag$_6$ and Ag$_{10}$ building blocks.}
\end{figure}

\begin{figure}
    \begin{center}
        \includegraphics[width=8.6cm]{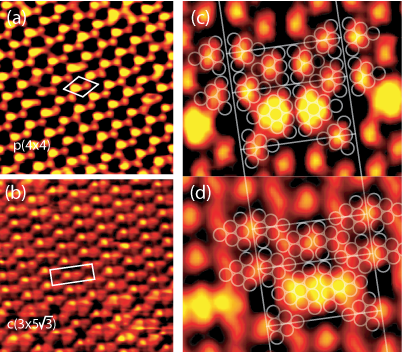}
    \end{center}
    \caption{\label{FigAg6TrianglesFccAndHcp}(Color online) Building block structures imaged
    with special tip states. (a) and (b) Imaging conditions which let the hcp and fcc Ag$_6$ triangles
    appear
    differently. (a) $100 \times 100$~{\AA}$^2$, 0.54~nA, 9.8~mV. (b) $100 \times 100$~{\AA}$^2$,
    0.85~nA, 55~mV. (c) and (d) Images of the same area in (c) normal imaging mode and (d)
    oxygen imaging mode. $47 \times 37$~{\AA}$^2$. (c) 1.48~nA, 55~mV. (d) 1.49~nA, 55~mV.}
\end{figure}

\subsection{Ag$_6$/Ag$_{10}$ building block structures\label{SecBuildingBlockStructures}}

The STM images of the $p(4 \times 5 \sqrt{3})$rect, $c(3 \times 5
\sqrt{3})$rect, and $p(4 \times 4)$ phases in
Figs.~\ref{FigAllStructures}(b-d) are characterized by ordered
arrays of unresolved bright features. Higher resolution images are
depicted in Figs.~\ref{FigBuildingBlockStructures}(a), (c), and
(e). From these it is seen that the bright features contain atomic
sub-structure. In the case of the $c(3 \times 5 \sqrt{3})$rect
phase a particular tip state even allowed all Ag atoms in the
overlayer to be resolved (inset of panel (c)). For the $c(3 \times
5 \sqrt{3})$rect and $p(4 \times 4)$ phases the arrangement of the
silver and oxygen atoms has already been discussed in detail in
Ref.~\onlinecite{Schnadt2006}. In brief, both structures are built
from close-to identical elements of Ag$_6$ triangles with the Ag
atoms of a single Ag$_6$ all located either in fcc or hcp sites.
Each primitive unit cell contains one fcc triangle and one hcp
triangle. The site difference implies a slight asymmetry between
the two triangles of a dimer, which, at least for the $p(4 \times
4)$ phase, is by DFT shown to be further enhanced by small
rotations in converse directions for the fcc and hcp
triangles\cite{Schmid2006,Reichelt2007a} [cf.
Fig.\ref{FigTheoryBBStructures}]. Indeed, for certain tip states
fcc triangles and hcp triangles are imaged differently in STM, as
is illustrated in Figure~\ref{FigAg6TrianglesFccAndHcp}(a-b). On
the basis of the STM measurements we thus propose the structural
models depicted in Figures~\ref{FigBuildingBlockStructures}(b) and
(d). It is difficult to provide a detailed assignment for the
location of the oxygen atoms, which normally are not visible in
STM due to the depletion of the local density of states at the
Fermi level typical for oxygen (see, however, below for an example
in which some oxygen density is visible). For the $c(3 \times 5
\sqrt{3})$rect and $p(4 \times 4)$ phases the oxygen atom
positions were derived by evaluating a large variety of different
structures in DFT calculations.\cite{Schnadt2006,Schmid2006} The
positions of the oxygen atoms in the thus derived structural
models have been included in the models shown in
Fig.~\ref{FigBuildingBlockStructures}(b) and (d), which were
suggested on the basis of the experimental STM results.
Figures~\ref{FigTheoryBBStructures}(a) and (b) show the structural
models emerging from the calculations. Particulary noteworthy are
the converse rotations of the Ag$_6$ triangles in the $p(4 \times
4)$ structure already mentioned above. The STM simulations, which
correspond to the calculated structures, are shown in panels (d)
and (e) and are found to be in good agreement with the
experimental STM results in
Figure~\ref{FigBuildingBlockStructures}(a) and (c). The DFT
adsorption energies (eV/O) are provided in
Table~\ref{TableStoichiometryDensity}. They are seen to be the
same to within the uncertainty of the DFT results.

The arrangement of the Ag atoms in the $p(4 \times 5
\sqrt{3})$rect structure in
Figure~\ref{FigBuildingBlockStructures}(e) can be understood from
an extension of the Ag$_6$ building block model for the $c(3
\times 5 \sqrt{3})$rect and $p(4 \times 4)$ phases. The larger
triangles constitute an additional Ag$_{10}$ building block, which
typically appears in dimer form, again with half of the silver
atoms in fcc and the other half in hcp sites. In the pure $p(4
\times 5 \sqrt{3})$rect phase the Ag$_6$ and Ag$_{10}$ dimers are
then arranged as depicted in
Fig.~\ref{FigBuildingBlockStructures}(f). It is noteworthy that
the $p(4 \times 5 \sqrt{3})$rect structure appears as a
mixture of the $p(4 \times 4)$ and $c(3 \times 5
\sqrt{3})$rect overlayers, with the voids between four Ag$_{10}$
and two Ag$_6$ triangles being $c(3 \times 5 \sqrt{3})$rect-like (dark
arrow in Fig.~\ref{FigBuildingBlockStructures}(f)) and those
between four Ag$_6$ and two Ag$_{10}$ being $p(4\times4)$-like
(light arrow).

As in the cases of the $p(4 \times 4)$ and $c(3
\times 5 \sqrt{3})$rect phases discussed above the oxygen atoms incorporated into
the $p(4 \times 5 \sqrt{3})$rect structure are normally not
visible in STM. Again, we carried out DFT calculations for a large number
of different arrangements of both oxygen and silver atoms to find
the most stable arrangement of surface atoms. This search also
included geometries which deviated from the experimentally derived
Ag$_{6}$/Ag$_{10}$ building block structure. Among the
investigated structures, the one displayed in
Figure~\ref{FigBuildingBlockStructures}(f) was found to be the
most stable one. The model is indeed based on the two different
building blocks suggested above on the basis of the STM results. The
oxygen atoms reside in the same sites as for the $p(4 \times 4)$
and $c(3 \times 5 \sqrt{3})$rect phases, a finding which lends
further credibility to the proposed model. The DFT oxygen
adsorption energy (eV/O) (cf.
Table~\ref{TableStoichiometryDensity}) is seen to be the same, to
within the DFT uncertainty, as that of the other two Ag$_6$
building block structures.

Fig.~\ref{FigTheoryBBStructures} contains the final structural
models for all three Ag$_6$/Ag$_{10}$ phases in panels (a)-(c).
The Figure also shows that the Tersoff-Hamann (TH) STM simulations
are in good agreement with the STM results. In particular, the TH
STM simulations reproduce the experimentally observed high
apparent height of the silver atoms in the centers of the
Ag$_6$/Ag$_{10}$ triangles, which are characterized by a high
silver coordination. The lower-Ag-coordinated edge atoms and even
more so the corner atoms of the Ag triangles have a much lower
local density of states intensity and therefore appear lower than
the triangle center atoms. We also note that by slightly modifying
the simulated tip apex height and bias conditions  we are not only
able to reproduce the experimentally observed triangular
appearance of the building blocks shown in
Figs.~\ref{FigTheoryBBStructures}(d-f), but also the more
spherical appearance of the bright features in
Fig.~\ref{FigAllStructures}(b-d).

The above derived models can explain the frequently observed
alignment of defects of the $p(4 \times 5 \sqrt{3})$rect type
embedded in a $c(3 \times 5 \sqrt{3})$rect domain. An example is
shown in Fig.~\ref{FigAllStructures}(c). A comparison of the
structural models for the $p(4 \times 5 \sqrt{3})$rect and $c(3
\times 5 \sqrt{3})$rect phases shows that the local geometry of
the Ag$_6$ building blocks is very similar in both phases, and, in
particular, the distance between the Ag$_6$ triangles along the
[11$\bar{2}$] surface direction is the same for both structures.
Thus, an aligned line of Ag$_{10}$ triangles, which is the second
building block of the $p(4 \times 5 \sqrt{3})$rect structure, fits
very well into a domain of the $c(3 \times 5 \sqrt{3})$rect phase.
If now, in a thought experiment, a single Ag$_{10}$ triangle is
moved from an aligned line to an isolated position, the structural
mismatch between the Ag$_{10}$ and Ag$_6$ triangles will either
lead to a reduction of the number of oxygen atoms accommodated in
the structure or, alternatively, to less favorable Ag-O bond
geometries. Either possibility will be accompanied by an energetic
cost, which, deeming from the experimental results, is not fully
compensated by the gain in configurational entropy.

In Table~\ref{TableStoichiometryDensity} we compare the
stoichiometries and top layer silver and oxygen atom densities of
the three Ag$_6$/Ag$_{10}$ building block structures. Both the
$p(4 \times 4)$ and $c(3 \times 5 \sqrt{3})$rect phases have a
Ag$_2$O stoichiometry, but the $c(3 \times 5 \sqrt{3})$rect has
slightly higher Ag and O atom densities than the $p(4 \times 4)$.
Judging from the stoichiometry the $p(4 \times 5 \sqrt{3})$rect
phase seems to be more oxygen-deficient; however, the absolute
oxygen density is the same for the $p(4 \times 4)$ and $p(4 \times
5 \sqrt{3})$rect structures, while the Ag density is increased for
the $p(4 \times 5 \sqrt{3})$rect and reaches the same level as in
the $c(3 \times 5 \sqrt{3})$rect. It is thus not correct to view
the $p(4 \times 5 \sqrt{3})$rect phase as a decomposition phase of
the $p(4 \times 4)$, as suggested by Carlisle \textit{et
al.}\cite{Carlisle2000b} While this characterization applies to
the $p(4 \times 5 \sqrt{3})$rect in relationship to the $c(3
\times 5 \sqrt{3})$rect phase, the $p(4 \times 5 \sqrt{3})$rect
structures rather corresponds to a Ag-rich reconstruction of the
$p(4 \times 4)$ structure.

\begin{figure}
    \begin{center}
        \includegraphics[width=8.6cm]{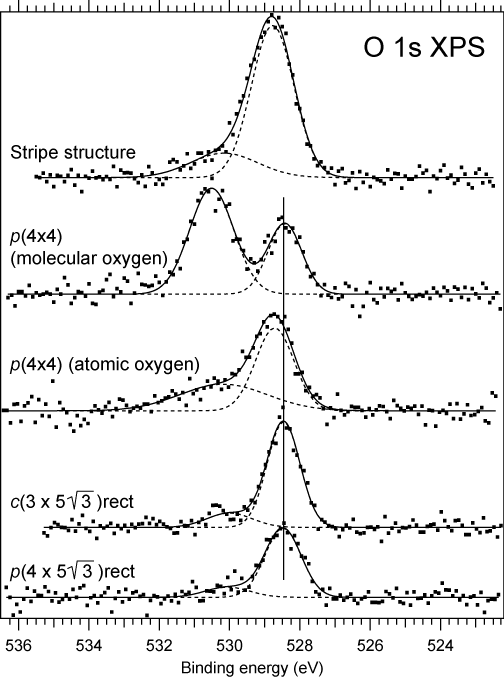}
    \end{center}
    \caption{\label{FigO1s}O~1s x-ray photoelectron spectra of the indicated preparations. The specifications "molecular oxygen" and "atomic oxygen"
    refer to which preparation method was used (cf. Table~\protect\ref{TableConditions}).}
\end{figure}

From the similar geometrical arrangements of the silver and oxygen
atoms in the three Ag$_6$/Ag$_{10}$ building block structures as
revealed in the STM images it seems likely that the electronic and
chemical characteristics of these phases should be very similar.
This is corroborated by the O~1s x-ray photoelectron spectra
depicted in Fig.~\ref{FigO1s}. Within the measuremental
uncertainty of 0.1~eV the O~1s binding energies are the same for
all three structures (528.5~eV) and they agree quite well with
previous measurements on the $p(4 \times 4)$
phase\cite{Campbell1985} (528.1~eV), although they are about
0.4~eV higher in energy (they are more similar to the recent
measurement by Reichelt \textit{et al.},\cite{Reichelt2007} who
report a binding energy of 528.3~eV). However, the O~1s peak
observed for the $p(4 \times 4)$ phase prepared from atomic oxygen
is located at 528.7~eV, and this upwards shift in energy we cannot
explain at present. It is further noted that all spectra contain a
high energy component at energies between 530.1 and 530.5~eV. This
peak is particularly pronounced for the $p(4 \times 4)$ phase
prepared using molecular oxygen, i.e., at an elevated pressure.
Previously, O~1s features at high binding energy have been
assigned to both
carbonates\cite{Rehren1991,Bukhtiyarov1999,Reichelt2007}, a
reactive oxygen species or chemisorbed
oxygen\cite{Bare1995,Bukhtiyarov1999,Reichelt2007}, and -- for the
Ag(110), Ag(210), and Ag(111) surfaces as well as polycrystalline
silver foil -- to subsurface or bulk-dissolved
oxygen.\cite{Rehren1991,PawelaCrew1995,Bao1996,Savio2006,Klust2006}
Considering that contaminations released from the reactor walls
are common in high pressure experiments such as used in the
molecular oxygen preparation of the $p(4 \times 4)$ phase (10~Torr
in the present case) it seems most likely that a large part of the
peak is related to carbonates. On the basis of the available data
we cannot, however, provide a conclusive assignment of this high
binding energy O~1s peak. Indeed, it is fully possible that this
peak contains contributions from several different oxygen species.

Finally it is noted, that it is sometimes possible to directly
image some of the surface oxygen density in the STM images (see,
e.g., also Ref.~\onlinecite{Merte2009}). As stated above, for
normal STM tunneling conditions oxygen adsorbed on Ag is not
easily identified in the STM images, but, as demonstrated in
Figure~\ref{FigAg6TrianglesFccAndHcp}(d) for a mixed $c(3 \times
\sqrt{5})$rect/ $p(4 \times \sqrt{5})$rect surface, rare tip
states seem to enhance those oxygen atoms which are coordinated by
four top layer silver atoms. The oxygen atoms are identified from
a comparison to the STM image in normal imaging mode in
Figure~\ref{FigAg6TrianglesFccAndHcp}(c) and the structural models
in Figure~\ref{FigBuildingBlockStructures}(d-f). In this special
tip state image the silver atoms of the smaller Ag$_6$ triangles
are not imaged at all, while the centers of the Ag$_{10}$
triangles appear nearly as bright as the fourfold-coordinated
oxygen atoms. The threefold-coordinated oxygens, however, remain
invisible even for the special tip state.

\begin{table*}
    \caption{\label{TableStoichiometryDensity}Stoichiometry, top layer Ag and O atom densities, and DFT adsorption energies $E_{ads}$
    of the
    here described O/Ag overlayers. The data for the $c(4 \times 8)$ structure are tentative, since the geometry model presented
    in the text so far has not been confirmed by further experimental and/or theoretical methods. $E_{ads}$ is defined
    as in Reference~\onlinecite{Michaelides2005}.}
    \begin{center}
    \begin{ruledtabular}
    \begin{tabular}[t]{ccccc}
                                     & Stoichiometry     & Ag density                 & O density                   & $E_{ads}$ (eV/O) \\
        \hline
        $p(4 \times 4)$              & Ag$_2$O           & 0.104 \AA$^{-2}$ (0.75 ML) & 0.052 \AA$^{-2}$ (0.375 ML) & 0.46 \\
        $c(3 \times 5 \sqrt{3})$rect & Ag$_2$O           & 0.111 \AA$^{-2}$ (0.80 ML) & 0.055 \AA$^{-2}$ (0.40 ML)  & 0.39 \\
        $p(4 \times 5 \sqrt{3})$rect & Ag$_2$O$_{0.94}$  & 0.111 \AA$^{-2}$ (0.80 ML) & 0.052 \AA$^{-2}$ (0.375 ML) & 0.42 \\
        $c(4 \times 8)$              & Ag$_2$O$_{1.6}$   & 0.043 \AA$^{-2}$ (0.31 ML) & 0.035 \AA$^{-2}$ (0.25 ML)  &  \\
    \end{tabular}
    \end{ruledtabular}
    \end{center}
\end{table*}

\begin{figure*}
    \begin{center}
        \includegraphics[width=17.8cm]{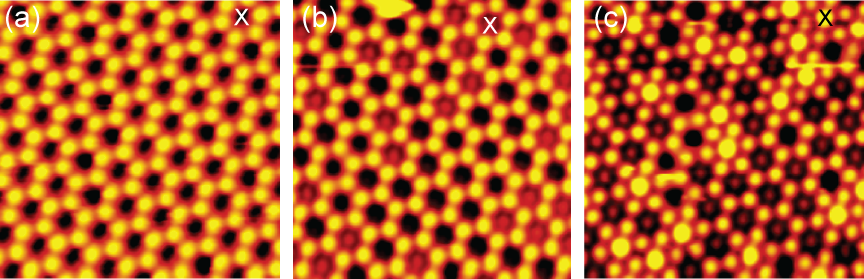}
    \end{center}
    \caption{\label{Figp4x4Fillings}(Color online) The $p(4 \times 4)$ phase as a host--guest architecture.
    The images were recorded on a sample prepared using atomic oxygen.
    (a) $90 \mbox{ \AA} \times 90 \mbox{ \AA}$, -0.35~nA, -15.9~mV.
    (b) $90 \mbox{ \AA} \times 90 \mbox{ \AA}$, -0.43~nA, -1030.3~mV.
    (c) $90 \mbox{ \AA} \times 90 \mbox{ \AA}$, -0.44~nA, -2024.2~mV.
    (a) and (c) were recorded at the same time (forward and backwards movements of the tip). (b) was recorded immediately
    prior to (a) and (c), and it was also accompanied by a simultaneous recording at low bias (-33~V). The latter image
    showed the same characteristics as the image in panel (a). The sample positions differ slightly between (a)/(c) and
    (b) due to thermal drift; the spatial relationship can, however, be inferred from the pattern of the hexagon
    fillings. Identical absolute positions have been marked by a cross.
}
\end{figure*}

\subsection{\label{SecHostGuest}The $p(4 \times 4)$ phase as a host-guest structure}

In Figure~\ref{Figp4x4Fillings} STM images of the $p(4 \times 4)$
phase are shown, which were recorded at varying bias voltages. As
shown in panel (a), at the lowest biases of approximately
$-15$~meV the STM image exhibits the typical honeycomb appearance
with the structural voids displayed as depressions (cf. the atomic
model in Fig.~\ref{FigBuildingBlockStructures}(b)). When the bias
is increased to about $-1000$~mV a fraction of the voids appears
filled (panel (b)). At the highest biases used, around $-2000$~mV,
the fillings already visible at the intermediate bias are now the
brightest features in the STM image, while other voids display
fillings with a smaller apparent height in the image [panel (c)].
Only a very minor fraction of voids seem to be unfilled.

We frequently observed the presence of fillings in the voids of
the $p(4 \times 4)$ structure; indeed, it is likely that they are
an inherent feature of the $p(4 \times 4)$ phase. The fillings are
most easily imaged at elevated biases of both tunneling
polarities. At low biases they normally remain unobserved, and,
thus, their absence in STM images recorded at low biases does not
imply their absence from the surface. The fillings occur in
preparations from both atomic and molecular oxygen, but we cannot
clarify here whether they exist for preparations from NO$_2$, as
well. Interestingly, we have not seen any such filling for the
$p(4 \times 5 \sqrt{3})$rect phase, although it -- in contrast to
the $c(3 \times 5 \sqrt{3})$rect structure -- contains the same
kind of voids as visible from the structural models in
Fig.~\ref{FigBuildingBlockStructures}.

We suggest that the fillings are related to adsorbate species
in the voids of the $p(4 \times 4)$ phase. Since we observed such fillings
both directly after preparation of the surface by either atomic or molecular oxygen
as well as after measurements for a time span of several hours, it seems
unlikely that rest gas contamination can be made responsible for the
fillings. Nonetheless, at present we cannot specify the chemical
nature of the fillings and we can just note that the observation
of the fillings suggests an interesting host-guest character
of the $p(4 \times 4)$ phase,
which would deserve a more systematic study.

\subsection{\label{Secc4x8stripe}The $c(4 \times 8)$ and stripe structure}

Finally we would like to discuss two additional structures not
observed previously. These are the $c(4 \times 8)$ and stripe
phases of Figure~\ref{FigAllStructures}(e-f). In both cases it is
difficult to provide the exact details of the atomic scale
structures. The reasons are that for the $c(4 \times 8)$ structure
we have so far not been able to find a recipe for preparing a
single phase $c(4 \times 8)$ surface, and for the stripe structure
there exists a larger manifold of similar, co-existing structures.
The difficulty in preparing surfaces covered by a single coherent
structure have prevented the use of averaging techniques such as
LEED, which would have resulted in useful additional information.
In the following we therefore limit ourselves to primarily
describing the characteristics of the structures to the extent
they could be extracted from the STM data.

\begin{figure}
    \begin{center}
        \includegraphics[width=8.6cm]{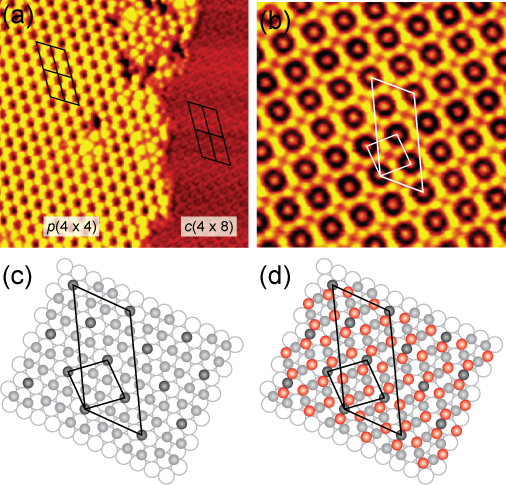}
    \end{center}
    \caption{\label{Figc4x8}(Color online) STM images and geometrical
    models of the $c(4 \times 8)$ phase. The colors are as in
    Figure~\protect\ref{FigBuildingBlockStructures}, with the
    addition that the darkest gray sphere represent the Ag atoms
    which appear isolated in the STM images.
    (a) $200 \mbox{ \AA} \times 200 \mbox{ \AA}$, -0.40~nA, -185.8~mV.
    The $c(4 \times 8)$ is here found to occur simultaneously with
    the $p(4 \times 4)$ phase. The large parallelograms depict arrays of
    four conventional unit cells of the $c(4 \times 8)$ structure.
    The superposition of the parallelogram onto the $p(4 \times
    4)$ phase confirms the $(4 \times 8)$ assignment, since
    each unit cell is seen to comprise two $p(4 \times 4)$ unit
    cells.
    (b) $60 \mbox{ \AA} \times 60 \mbox{ \AA}$, -0.34~nV, -109.6~mV.
    (c) Tentative geometry of the Ag atomes in the $c(4 \times 8)$ structure.
    The parallelograms in (b) to (d) show both the
    conventional and primitive unit cells.
}
\end{figure}

Starting with the $c(4 \times 8)$ structure, its symmetry
assignment was derived from comparing its unit cell with those of
other phases appearing in the same STM image. Such a comparison is
shown in Fig.~\ref{Figc4x8}(a). The correctness of the assignment
is supported by the good agreement between the FFT of the STM
image and the LEED simulation based on the $c(4 \times 8)$ unit
cell in Figure~\ref{FigAllStructures}(e). Since oxygen typically
is invisible in the STM images it is assumed that the bright
features in the more highly resolved STM images of the $c(4 \times
8)$ structure correspond to silver atoms, cf.
Fig.~\ref{Figc4x8}(b). This assignment leads us to suggest the
tentative top layer Ag atom geometry depicted in
Fig.~\ref{Figc4x8}(c). It is the only geometry for which the
bright features in the experimental image can be brought into
coincidence with lattice sites. In this model, 20\% of the top
layer silver atoms reside in bridging sites, while the remaining
silver atoms are found in both hcp and fcc hollow sites.

The ring-formed depressions around the Ag atoms in bridging sites
indicate that the Ag atoms are surrounded by oxygen atoms. The
placement of four oxygen atoms in the dark ring leads to the
detailed geometry shown in Figure~\ref{Figc4x8}(d). In this
geometry the coordination of the silver atoms by other top layer
silver atoms is much lower than in the building block structures
described in section \ref{SecBuildingBlockStructures}. In these
structures silver atoms with a low silver coordination such as the
corner atoms of the Ag$_6$/Ag$_{10}$ triangles were imaged with a
reduced apparent height as compared to the fully
silver-coordinated silver atoms in the centers of the triangles.
As seen clearly in Figure~\ref{Figc4x8}(a), also the $c(4 \times
8)$ phase is characterized by a much smaller apparent height than
the brightest features of the $p(4 \times 4)$ phase. We take this
as an indication that the $c(4 \times 8)$ structure indeed is more
highly oxidized than any of the building block structures.

The suggested structure for the $c(4 \times 8)$ phase possesses a
Ag$_5$O$_4$ (Ag$_2$O$_{1.6}$) stoichiometry (cf.
Table~\ref{TableStoichiometryDensity}). A very similar structure
was observed previously for a Pd$_5$O$_4$ surface
oxide.\cite{Lundgren2002} A major difference between the present
and the Pd oxide case is, however, the full commensurability of
the Ag$_5$O$_4$ unit cell, while the Pd$_5$O$_4$ is incommensurate
in one surface direction. It is seen that the proposed
stoichiometry for the $c(4 \times 8)$ is more reminiscent of AgO
than Ag$_2$O and that the density of Ag and O atoms is much lower
in the $c(4 \times 8)$ as compared to the Ag$_6$/Ag$_{10}$
building block structures. This difference in stoichiometry and Ag
and O densities in comparison to the $p(4 \times 4)$, $c(3 \times
5\sqrt{3})$, and $p(4 \times 5\sqrt{3})$ phases is surprising in
view of the coexistence of these structures such as those in
Figure~\ref{FigStructuresAtthesametime}. At present we cannot
satisfactorily explain this finding.

\begin{figure}
    \begin{center}
        \includegraphics[width=8.6cm]{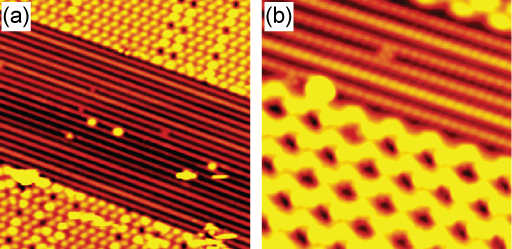}
    \end{center}
    \caption{\label{FigStripes}(Color online) STM images of the
    stripe phase.
    (a) $200 \mbox{ \AA} \times 200 \mbox{ \AA}$, -0.41~nA, -25.0~mV.
    (b) $75 \mbox{ \AA} \times 75 \mbox{ \AA}$, -0.47~nA, 58~mV.
}
\end{figure}

Now turning to the stripe phase, we would first like to again
emphasize that there exists a larger number of co-existing
structures which appear as stripes in the STM measurements.
Probably these structures have similar, but not exactly equal
atomic geometries. That the geometries indeed differ between
different domains of the stripe phase is quite obvious from a
comparison of the displayed domains in
Figs.~\ref{FigAllStructures}(f) and \ref{FigStripes}. The common
denominator of the domains is the occurrence of stripes along
\{1$\bar{1}$0\}, but the arrangement and apparent height of the
stripes are not always the same.

While it was difficult in general to obtain high resolution STM
images on extended domains of stripe character, this was sometimes
possible on smaller domains embedded in other structures.
Figure~\ref{FigStripes} provides examples of such high-resolution
STM images. We tentatively assume that the protruding features of
these images are silver atoms. Along the \{1$\bar{1}$0\}
directions the distance between these features corresponds to the
Ag(111) nearest-neighbor distance $b=2.89$~{\AA}.  In the
perpendicular direction the distances of the rows is, however, not
a multiple of $\sqrt{3} b$, as might have been expected. By
comparison to the surrounding surface covered by domains of the
$p(4 \times 4)$, $c(3 \times 5 \sqrt{3})$rect, and $p(4 \times 5
\sqrt{3})$rect phases it is estimated that the repetitive unit
rather comprises four stripes and that this unit has a width of $4
\sqrt{3} b$. In such a unit cell it is not possible to place all
the silver atoms in the same lattice site. The differing
brightness of adjacent stripes supports this conclusion.

The position of the oxygen atoms in the stripe phase cannot be
specified in any detail from an analysis of the STM results. It
seems likely that oxygen atoms reside in between the bright
stripes. The XPS data in Figure~\ref{FigO1s} indicate that the
oxygen content is considerably higher than in the $p(4 \times 4)$
and $c(3 \times 5 \sqrt{3})$rect phases. Similar to the $c(4
\times 8)$ structure and the corner atoms of the triangles in the
building block structures, the stripes are imaged at a smaller
apparent height compared to that of the adjacent domains of
primarily $c(3 \times 5 \sqrt{3})$rect symmetry in panel (a) of
Figure~\ref{FigStripes} and primarily $p(4 \times 4)$ in panel
(b). This suggests that the silver coordination of the Ag atoms of
the stripe phase is smaller than that of the Ag atoms in the
centers of the triangles of the building block structures, which
indicates a higher oxygen coordination and thus a higher oxygen
content of the stripe phase. However, at present, we refrain from
a more quantitative analysis, partly since it is unclear to what
extent the XPS data represent fully covered surfaces and partly
since the STM data cannot provide any more detailed information.
It is noted, though, that metal/oxygen structure with quite a high
oxygen content and exhibiting a stripe appearance have been
observed on other metal surfaces such as Pt(110).\cite{Helveg2007}

\section{Conclusions}

We have shown from an interplay of STM and XPS experiments and DFT
calculations that the oxidation of the Ag(111) surface may be
structurally more complex than previously anticipated. Depending
on the preparation conditions -- oxidation agent, dose, and
pressure, exposure time, and sample temperature during and after
exposure -- a large variety of different structures are found,
essentially all of which may coexist. Among these phases we have
identified a number of structures, which are formed from
Ag$_6$/Ag$_{10}$ fundamental building blocks. This group of
structures comprises both the renowned $p(4 \times 4)$ surface as
well as two phases of $c(3 \times 5 \sqrt{3})$rect and $p(4 \times
5 \sqrt{3})$rect symmetry. The $p(4 \times 5 \sqrt{3})$rect
structure had been observed previously,\cite{Carlisle2000b} but we
have devised a structural model which is in line with the
geometries for the $p(4 \times 4)$ (Refs. \onlinecite{Schnadt2006}
and \onlinecite{Schmid2006}) and $c(3 \times 5 \sqrt{3})$rect
phases (Ref. \onlinecite{Schnadt2006}) proposed recently and
discussed in more detail here. In addition, we have observed two
additional structures of $c(4 \times 8)$ and stripe character.

The present results suggest that a complex coexistence of oxide
and oxygen adsorbate overlayers with varying contents of oxygen
may form under the conditions of industrial oxidation catalysis.
This is in contrast to the prevalent perception that the $p(4
\times 4)$ phase alone represents an adequate model for the
surface under such conditions. It is conceivable that the
complexity of the phase diagram has a profound influence on the
dynamics of the catalytic process. Hence, although an increased
understanding of the structure of the oxidized silver surface is
presently being obtained as demonstrated in this and other
studies,\cite{Schnadt2006,Schmid2006,Klust2007,Reicho2007,Reichelt2007,Reichelt2007a}
we are still far from a detailed understanding of the oxidized
silver surface under reaction conditions. It appears likely that
the structure of the silver oxide surface may vary in response to
changes in the gas composition and the chemical potential of the
reaction intermediates.

\acknowledgments

The competent assistance of the staff of the mechanical workshop
of the Department of Physics and Astronomy in Aarhus is gratefully
acknowledged. JS wishes to thank the European Commission for
funding through a Marie Curie Intra-European Fellowship. AM is
grateful to the Alexander von Humboldt foundation for partial
support of this work and the European Science Foundation and EPSRC
for a European Young Investigator Award (EURYI). Some of the
calculations performed here were made possible by AM's membership
of the UK's HPC Materials Chemistry Consortium, which is funded by
EPSRC (EP/F067496).


\end{document}